\documentstyle[12pt,aaspp4]{article}

\begin{document}

\title{Predictions for The Very Early Afterglow and The Optical Flash}
\author{Re'em Sari$^1$ and Tsvi Piran$^2$ \\
$^1$ Theoretical Astrophysics 130-33, California Institute of Technology,
Pasadena, CA 91125, USA \\
$^2$ Racah Institute of Physics, The Hebrew University, Jerusalem 91904, 
Israel and Department of Physics, Columbia University, New York, NY 10027, USA}

\begin{abstract}
According to the internal-external shocks model for $\gamma $-ray
bursts (GRBs), the GRB is produced by internal shocks within a
relativistic flow while the afterglow is produced by external shocks
with the ISM. We explore the early afterglow emission. For short GRBs
the peak of the afterglow will be delayed, typically, by few dozens of
seconds after the burst. For long GRBs the early afterglow emission
will overlap the GRB signal. We calculate the expected spectrum and
the light curves of the early afterglow in the optical, X-ray and
$\gamma $-ray bands. These characteristics provide a way to
discriminate between late internal shocks emission (part of the GRB)
and the early afterglow signal.  If such a delayed emission, with the
characteristics of the early afterglow, will be detected it can be
used both to prove the internal shock scenario as producing the GRB,
as well as to measure the initial Lorentz factor of the relativistic
flow. The reverse shock, at its peak, contains energy
which is comparable to that of the GRB itself, but has a much lower
temperature than that of the forward shock so it radiates at considerably
lower frequencies. The reverse shock dominates the early optical emission,
and an optical flash brighter than 15th magnitude, is expected together 
with the forward shock peak at x-rays or $\gamma$-rays.
If this optical flash is not observed, strong limitations can be put 
on the baryonic contents of the relativistic shell deriving the GRBs, 
leading to a magnetically dominated energy density.
\end{abstract}


\section{Introduction}

The original fireball model was invoked to explain the Gamma-Ray Bursts
(GRBs) phenomena. Extreme relativistic motion, with Lorentz factor $\gamma
>100$ is necessary to avoid the attenuation of hard $\gamma $-rays due to
pair production. Such extreme relativistic bulk motion is not seen anywhere
else in astrophysics. This makes the GRBs a unique and extreme phenomena.
Within the fireball model the observed GRB and the subsequent afterglow
all emerge from shocked regions in which the relativistic flow is slowed
down. We don't see directly the ``inner engine'' which is the source
of the whole phenomenon. It is therefore, of the utmost importance to 
obtain as much information as possible on the nature of this flow 
as this would provide us with some of the best clue on what is producing
GRBs.

The afterglow, that was discovered more than a year ago, has
revolutionized GRB Astronomy. It proved the cosmological origin of the
bursts. The observations, which fit the fireball theory fairly well
are considered as a confirmation of the fireball model. According to
this model the afterglow is produced by synchrotron radiation produced
when the fireball decelerates as it collides with the surrounding
medium.

However, the current afterglow observations, which detect radiation
from several hours after the burst onwards, do not provide a
verification of the initial extreme relativistic motion. Several hours
after the burst the Lorentz factor is less than $\sim
10$. Furthermore, at this stage it is independent of the initial
Lorentz factor. These observation do not provide any information on
the initial extreme conditions which are believed to produce the burst
itself.

It was recently shown (Sari \& Piran, 1997, Fenimore, Madras \&
Nayakshin 1996) that the burst itself cannot be efficiently produced
by external shocks, and internal shocks must occur. This has lead to
the internal-external scenario. The GRB is produced by internal shocks
while the afterglow is produced by external shocks. 
Additionally there
are some observational evidence in favor of the internal-external
picture. First, the fact that afterglows are not scaled directly to
the GRB suggest that the two are not produced by the same
phenomenon. Second, while most GRBs show very irregular time structure
and are highly variable all afterglow observed so far show smooth
power law decay with minimal or no variability. Still this evidence is
so far somewhat inconclusive. In view of the importance of its
implications we should search for an additional proof.  We suggest
here that observation of the early afterglow could provide us with a
verification of this picture.

In the internal
shocks GRB the time scale of the bursts and its overall temporal
structure follow to a large extend the temporal behavior of the source
which generates the relativistic flow and powers the GRB (Kobayashi,
Piran \& Sari, 1997). A fast shell, with a Lorentz factor $>2\gamma $,
will catch up with a slower shell of Lorentz factor $\gamma $, that
was emitted $\delta t$ earlier, at a radius of $R\sim 2 \gamma
^{2}c\delta t$.  The observed time for this collision will be
therefore $R/2 \gamma ^{2}\sim \delta t$. The fact that the Lorentz
factor cancels out shows that the observed temporal structure of the
burst cannot provide any information on the initial Lorentz factor in
which the shell was injected.

The initial Lorentz factor is a crucial ingredient for constraining models
of the source itself. The initial Lorentz factor specifies how ``clean''
the fireball is as the baryonic load is $M=E/\gamma _{0}$. A very high
Lorentz factor would indicate a very low baryonic load which would indicate
some sort of electromagnetic acceleration or even Poynting flux flow. More
moderate Lorentz factors could more easily allow for the usual hydrodynamic
models. The previous discussion shows that we cannot infer on the initial
Lorentz facer from the observed temporal structure in GRBs. Unfortunately
the spectrum of the GRBs  can provide only a lower limit to this  this   
Lorentz factor. This lower limit of $\sim 100$ is given by the appearance of 
high energy photons, which would have produces pairs is the Lorentz factor
was low (Fenimore, Epstein,  \& Ho, 1993; Woods \& Loeb, 1995;
Piran, 1997).
In the internal shock scenario the observed spectrum depends on the
Lorentz factor only via the blue shift. However, the frequency in the local
frame is highly unknown since it depends on many poorly known parameters
such as the fraction of energy given to the electrons, the magnetic field
and the relative Lorentz factor between shells. Therefore the spectrum can
not teach us much about the initial Lorentz factor.

Furthermore, the basic mechanism by which the burst is produced, internal or
external shocks, must be understood before a reliable source model can
be given.  External shocks could be produced by a single short
explosion. Internal shocks require, however, a long and highly
variable wind. The inner engine should operate for a long time - as
long as the duration of the burst. We must know whether the source
operates for a millisecond or for tens or even hundreds or thousand of
seconds.

Mochkovitch, Maiti \& Marques (1995) and Kobayashi, Piran \& Sari, (1997)
have shown that only a fraction of the total energy of the
relativistic flow could be radiated away by the internal shocks. This
means that an ample amount of energy is left in the flow and a
significant fraction of it can be emitted by the early afterglow. 

GRBs are among the most luminous objects in the universe. They produce a huge
fluence, mostly in $\gamma$-rays. If this fluence was released optically, a 
flash of 5th magnitude would have been produced.
A magnitude of 5 is by far stronger
than current observational upper limits on early optical emission. 
In fact, even a small
fraction of this will be easily observed.  It is therefore of importance
to calculate any residual emission in the optical band
(Sari \& Piran 1999).

We explore in this paper the expected prompt (early afterglow)
multi wavelength signal. We show that this initial afterglow signal
could, when it be measured, provide us with information on the initial
Lorentz factor and at least indirectly hint on the nature of the
relativistic flow. These could provide some important clues on the
nature of the ``inner engine'' that powers GRBs.  The physical model
of the afterglow is synchrotron emission from relativistic electrons
that are being continuously accelerated by an ongoing shock with the
surrounding medium. We consider both the emission due to the
hot shocked surrounding medium and from the reverse shock that is
propagating into the shell. The spectral characteristics of the
synchrotron emission process are unique while the light curve depends
on the hydrodynamic evolution, which is more model dependent. We
begin, therefore (in section 2) exploring the broad band
spectrum due to the forward shock. 
Then we turn (in section 3) to discuss the possible light
curves in several frequency regimes. In section 4 we show how future
observations of the early afterglow can be used to estimate the
initial Lorentz factor. We show that a detection of a delay between
the GRB and its afterglow as well as observation of the characteristic
frequency in the early afterglow can finally provide a strong evidence
for the internal shocks mechanism.
In section 5 we calculate the optical emission including that expected from
the reverse shock. 

\section{Synchrotron Spectrum of Relativistic Electrons.}

The synchrotron spectrum from relativistic electrons that are
continuously accelerated into a power law energy distribution is
always given by four power law segments, separated by three critical
frequencies: $\nu _{sa}$ the self absorption frequency, $\nu _{c}$ the
cooling frequency and $\nu _{m}$ the characteristic synchrotron
frequency. The electrons could be cooling rapidly or slowly
and this would change the nature of the spectrum (Sari,
Piran \& Narayan 1998). 
As we show later, only fast cooling is relevant during the
early stages of the forward shock (except perhaps the first second). 
We consider, in the following, only this fast cooling regime.

The spectrum of fast cooling electrons is described by
four power laws: (i) For $\nu <\nu _{sa}$ self absorption is
important and $F_{\nu }\propto \nu ^{2}$. (ii) For $\nu _{sa}<\nu <\nu
_{c}$ we have the synchrotron low energy tail $F_{\nu }\propto \nu
^{{1/3}}$. (iii) For $\nu _{c}<\nu <\nu _{m}$ we have the electron
cooling slope $F_{\nu }\propto \nu ^{-1/2}$. (iv) For $\nu >\nu _{m}$
the spectrum depends on the electron's distribution, $F_{\nu }\propto
\nu ^{-p/2}$, where $p$ is the index of the electron power law
distribution. This spectrum is plotted in figure 1. This figure is a
generalization of figure 1a of Sari, Piran and Narayan (1998) for
an arbitrary hydrodynamic evolution $\gamma (R)$.

\begin{figure}[tbp]
\begin{center}
\epsscale{1.} \plotone{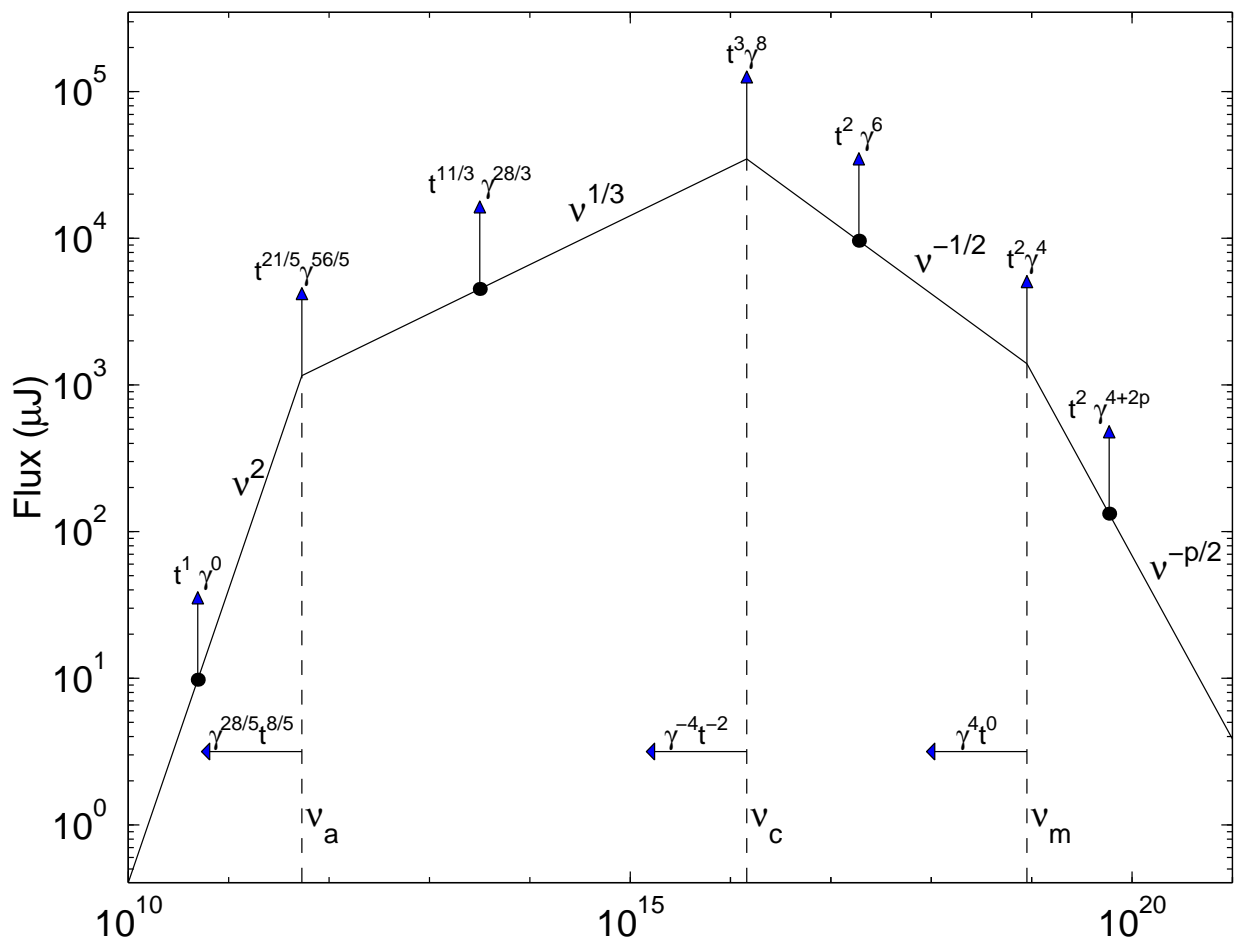}
\end{center}
\caption{ The synchrotron spectrum of power law injected electrons, with
cooling and self absorption. The spectrum is made out of four power laws
indicated in the figure. The flux at a given (fixed) frequency at each of
these segments changes with time as indicated above the up-arrows which
begin with a circle. The circle-less arrows, plotted at the break points
indicate the scaling of the flux at the break point. }
\label{fig1}
\end{figure}

Using the shock jump condition and assuming the electrons and magnetic field
acquire a fraction of $\epsilon _{e}$ and $\epsilon _{B}$ of equipartition,
we can describe all hydrodynamic and magnetic conditions behind the shock as
a function of the observed time $t=t_s$sec, the Lorentz factor $\gamma $ 
and the surrounding density $n_1$ cm$^{-3}$. 
The magnetic field $B$ and the typical electron Lorentz factor $\gamma _{e}$ 
are given by: 
\begin{equation}
B=4 \gamma \sqrt{2\pi \epsilon _{B}nm_{p}c^{2}},
\end{equation}
\begin{equation}
\gamma _{e}=610\epsilon _{e}\gamma .
\end{equation}
The typical synchrotron frequency of such an electron is 
\begin{equation}
\nu _{m}=1.1\times 10^{19}{\rm Hz}\left( \frac{\epsilon _{e}}{0.1}\right)
^{2}\left( \frac{\epsilon _{B}}{0.1}\right) ^{1/2}(\frac{\gamma}{300}%
)^{4}n_{1}^{1/2}.
\label{numf}
\end{equation}
Within the dynamical time of the system, the electrons are cooling down to
a Lorentz factor $\gamma _{c}$ where the total energy emitted at a time $t$
is comparable to the electron's energy: $\sigma _{T}c\gamma ^{2}\gamma
_{c}^{2}B^{2}t/6\pi =\gamma _{c}m_{e}c^{2}\gamma $. The cooling
frequency is the synchrotron
frequency $\nu _{c}$ of such an electron:
\begin{equation}
\nu _{c}=1.1\times 10^{17}{\rm Hz}\left( \frac{\epsilon _{B}}{0.1}\right)
^{-3/2}\left( \frac{\gamma }{300}\right) ^{-4}n_{1}^{-3/2}t_{s}^{-2},
\label{nucf}
\end{equation}
where throughout this paper we use $R\sim 2\gamma ^{2}t$, leaving
aside corrections of order unity\footnote{The numerical coefficient
chosen is a compromise between that suitable for the burst itself, and
that suitable for the deceleration phase.}.  One can see that for
typical parameters, the cooling frequency is lower than the typical
synchrotron frequency, except for a very short initial time (0.1
second for $\epsilon_e=\epsilon_B=0.1$ $\gamma_0=300$).

The flux
at $\nu_c$ is given by the number of radiating electrons, $4\pi
nR^{3}/3$ times the power of a single electron: 
\begin{equation}
F_{\nu ,\max }=220{\rm \mu J}D_{28}^{-2}\left( \frac{\epsilon _{B}}{0.1}%
\right) ^{1/2}\left( \frac{\gamma }{300}\right) ^{8}n_{1}^{3/2}t_{s}^{3}.
\end{equation}
\newline
Finally the self absorption frequency is given by the condition that the
optical depth is of order unity i.e. 
\begin{equation}
\label{nusa}
\nu _{sa}=220{\rm GHz}\left( \frac{\epsilon _{B}}{0.1}\right) ^{6/5}\left( 
\frac{\gamma }{300}\right) ^{28/5}n_{1}^{9/5}t_{s}^{8/5}.
\end{equation}
\newline
These scaling are all indicated in figure 1. Note that some of these
numbers involve high powers of $\gamma $ and $t$. Therefore, the
numerical coefficient given, can be considerably different from the
actual value with only a slight change of these parameters. Note that
when there are high powers of $t$, the numerical factor in the
approximation $t=R/2\gamma ^{2}c$ will also affect the numerical
result.

The above equation shows that the frequency range for the forward shock,
though depends strongly on the systems parameters, is most likely to be 
around the hard x-ray to $\gamma$-ray regime. This is more or less like
the observed GRB. The fraction of the electrons energy that is emitted 
in optical
bands is very small. We shall show later that the reverse shock emission
is at a considerably lower frequency, typically around the optical band.  
However we ignore the reverse shock emission at this stage, and turn to 
calculate the light curves produced by the forward shock.

\section{Light Curves}

While the spectrum is always described by the four broken power laws of
figure 1, the light curves depend on how the hydrodynamic conditions vary
with time. The temporal scaling within each of the spectral segments
appearing in figure 1 are given by the up, circle based arrows, when
substituting the scaling of $\gamma $ as a function of $t$. These scalings
depend on the exact form of the hydrodynamic evolution. Two shocks are
formed as the shell propagates into the surrounding material. A forward
shock accelerating and heating the surrounding material and a reverse shock
decelerating the shell (Rees \& M\'{e}sz\'{a}ros 1992, Katz 1994, Sari \&
Piran 1995). 

Consider a relativistic shell with an initial width in the observer
frame, $\Delta $, and an initial Lorentz factor, $\gamma _{0}$. Sari
and Piran (1995) have shown that there are four critical hydrodynamic
radii: $R_{s}\sim \Delta \gamma _{0}^{2}$ where the shell begin to
spread; $R_{\Delta }\sim (\Delta E/nm_{p}c^{2})^{1/4}$ where the
reverse shock crossed the entire shell; $R_{\gamma }\sim
(E/nm_{p}c^{2})^{1/3}\gamma _{0}^{-2/3}$ where a surrounding shocked
mass smaller by a factor of $\gamma _{0}$ from the shell's rest mass
($E/\gamma _{0}c^{2}$) was collected; $R_{N}\sim (E/nm_{p}c^{2}\Delta
)^{1/2}\gamma _{0}^{-2}$ where the reverse shock becomes relativistic.

We divide the different configurations according to the relative
``thickness'' of the relativistic shell. The question whether a shell should
be considered as thin or thick depends not only on its thickness, $\Delta $,
but also on its Lorentz factor $\gamma _{0}$. We will consider a shell thin
if $\Delta <(E/nm_{p}c^{2})^{1/3}\gamma _{0}^{-8/3}$. Shells satisfying $
\Delta >(E/nm_{p}c^{2})^{1/3}\gamma _{0}^{-8/3}$ are considered thick. 

For thin shells the corresponding transition radii are ordered as $
R_{s}<R_{\Delta }<R_{\gamma }<R_{N}$. As the shell expands it begins
to spread at $R_{s}$. For $R>R_{s}$ the width increases and this
causes $ R_{\Delta }$ to increase and $R_{N}$ to decrease in such a
way that $ R_{\Delta }=R_{\gamma }=R_{N}$. So if spreading occurs, by
the time when the reverse shock crosses the shell it is mildly
relativistic. The corresponding observed time scale of the early
afterglow is therefore $t_\gamma=R_{\gamma }/2\gamma _{0}^{2}$. This is longer
than the burst's duration, $\Delta /c$, so a separation between the
burst and the afterglow is expected (Sari 1997).

For thick shells, the order is the opposite $R_{N}<R_{\gamma
}<R_{\Delta }<R_{s}.$ The reverse shock becomes relativistic early on,
reducing the Lorentz factor of the shell as $\gamma \sim t^{-1/4}$
(Sari 1997). The radius $R_{\gamma }$ becomes unimportant and most of
the energy is extracted only at $R_{\Delta }$, with an observed
duration of order $\Delta /c$. The signals from the internal shocks
(the GRB) and from the early external shocks (the afterglow) from a
thick shell overlap. For thick shells, it might be difficult, 
therefore, to detect the smooth external shock component.

The final self-similar deceleration phase does not depend on the
thickness of the shell.  After most of the energy of the shell was
given to the surrounding (at $ R_{\gamma }$ for  thin shells  and
at $R_{\Delta }$ for  thick shells) the deceleration goes on as
$\gamma \sim t^{-3/8}$, in a self-similar manner.

To summarize, the hydrodynamic evolution can have two or three stages. 
In the first stage, the
ambient mass is too small to affect the system (the reverse shock is weak)
and the Lorentz factor is constant. In the last stage the deceleration 
is self similar with $\gamma \sim t^{-3/8}$, this stage lasts for months.
An intermediate stage of $\gamma \sim t^{-1/4}$ may occur for thick shells
only. Most of the energy is transfered to the surrounding material
at $t_\gamma$ for thin shells or $\Delta/c$ for thick shells.

If internal shocks give rise to the GRB then the observed duration of
the burst equals the initial width of the shell divided by $c$. Short
bursts correspond to thin shells and long bursts to thick ones. The
thickness of the shell in the internal shock scenario is directly
observed. Bursts of $20$ sec or longer are likely to belong to the
thick shell category. While bursts of duration smaller than $0.1$sec
are likely to belong to the thin shell category unless $\gamma _{0}$
is very large ($1500$ or larger). 
If internal
shocks are to produce the bursts, they must occur before the reverse
shock has crossed the shell.  Since the typical
collision radius for internal shocks is $ 2\delta \gamma _{0}^{2}$,
where $\delta $ is the separation between the shells,
one needs $2\delta \gamma _{0}^{2}<R_{\Delta }$. 
This is satisfied automatically for thin
shells, for which $2\delta \gamma _{0}^{2}<2\Delta\gamma_0^2=R_{s}<R_{\Delta
}$. However, an additional constraint: $2\delta \gamma
_{0}^{2}<(\Delta E/nm_{p}c^{2})^{1/4}$ arises for thick shells, and set
an upper limit to the initial Lorentz factor $\gamma_0$.

Equations \ref{numf}-\ref{nusa} show that the self absorbed flux 
always rises as $t^{1}$. This behavior is
independent of the hydrodynamic evolution and it is a general 
characteristic of fast cooling emission from the forward shock. 
Therefore, in principle, it can be used as a test of whether the
electrons are cooling rapidly or not. However, the self absorbed emission, 
which is relevant at radio frequencies, is very week in
the early afterglow. Detection of radio emission within a few seconds
of the burst is unlikely in the near future.  Therefore,
we will not discuss further the self absorbed frequencies in this
paper.

There are many possible light curves. This follows from the appearance
of numerous transition between different hydrodynamic evolutions and
between the four different spectral segments.  Similar to Sari,
Narayan and Piran (1998), we define the (frequency dependent) times
$t_c$ and $t_m$ as the time where the cooling and typical frequencies,
respectively, cross the observed frequency.  Different time ordering
of these transitions would lead to different light curves.  For thick
shells, there are two spectral related times $t_{c}<t_{m}$, as well as
two hydrodynamic transitions occurring at $R_N<R_\Delta$ which
corresponds to times $t_{N}<t_{\Delta }$.  There are therefore six
possible orderings and six corresponding light curves. However, as we
have mentioned earlier the initial afterglow signal from a thick shell
overlaps the GRB. It will be hard to detect the initial smooth signal
of the afterglow and to separate it from the complex internal shocks
signal. In addition, as we show later, thin shells can provide more
information on the initial Lorentz factor. We will therefore, consider 
in the rest of the paper only the
light curves produce by thin shells.

Thin shells are easier to analyze. Here we have two spectral related
times $ t_{c}<t_{m}$, but only one additional hydrodynamic time
$t_{\gamma }$. At $t_{\gamma }$ the flow changes from a constant
Lorentz factor into the self-similar decelerating phase. The thin
shell deceleration time, $ t_{\gamma }$, is given by
\begin{equation}
t_{\gamma }=R_{\gamma }/2\gamma _{0}^{2}c=\left( \frac{3E}{32\pi \gamma
_{0}^{8}nm_{p}c^{5}}\right) ^{1/3}  \label{tofgamma}
\end{equation}
There are only 3 possible light curves. Moreover, it will be
easier to distinguish between the GRB and the early afterglow emission from
thin shells, as there is a delay between the two.

The light curve is determined according to the time ordering
of the different time scales which varies from one observed frequency
to the other.  We consider, first, high frequencies that are above the
initial typical synchrotron frequency (the typical synchrotron
frequency with the initial Lorentz factor $\gamma _{0}$):
\begin{equation}
\label{highf}
\nu >1.1\times 10^{19}\left( \frac{\epsilon _{e}}{0.1}\right) ^{2}\left( 
\frac{\epsilon _{B}}{0.1}\right) ^{1/2}(\frac{\gamma _{0}}{300}%
)^{4}n_{1}^{1/2}.  
\end{equation}
The typical synchrotron frequency, $\nu _{m}$, depends only of $\gamma
$. Since $\gamma$ always decreases with time, then $\nu_m$ will
decrease with time as well. Therefore, if the observed frequency is
initially above the initial typical synchrotron frequency it will
remain so during the whole evolution. Consequently, the time $ t_{m}$ is
not defined for these high frequencies. The light curve in this high
frequency regime will be given by $t^{2}\gamma ^{4+2p}$ throughout the
hydrodynamic evolution (see  figure 2a). The light curve rises initially,
when $\gamma$ is a constant and then it decreases sharply when $\gamma$
begins to decline.

\begin{figure}
\begin{center}
\epsscale{.8} 
\plotone{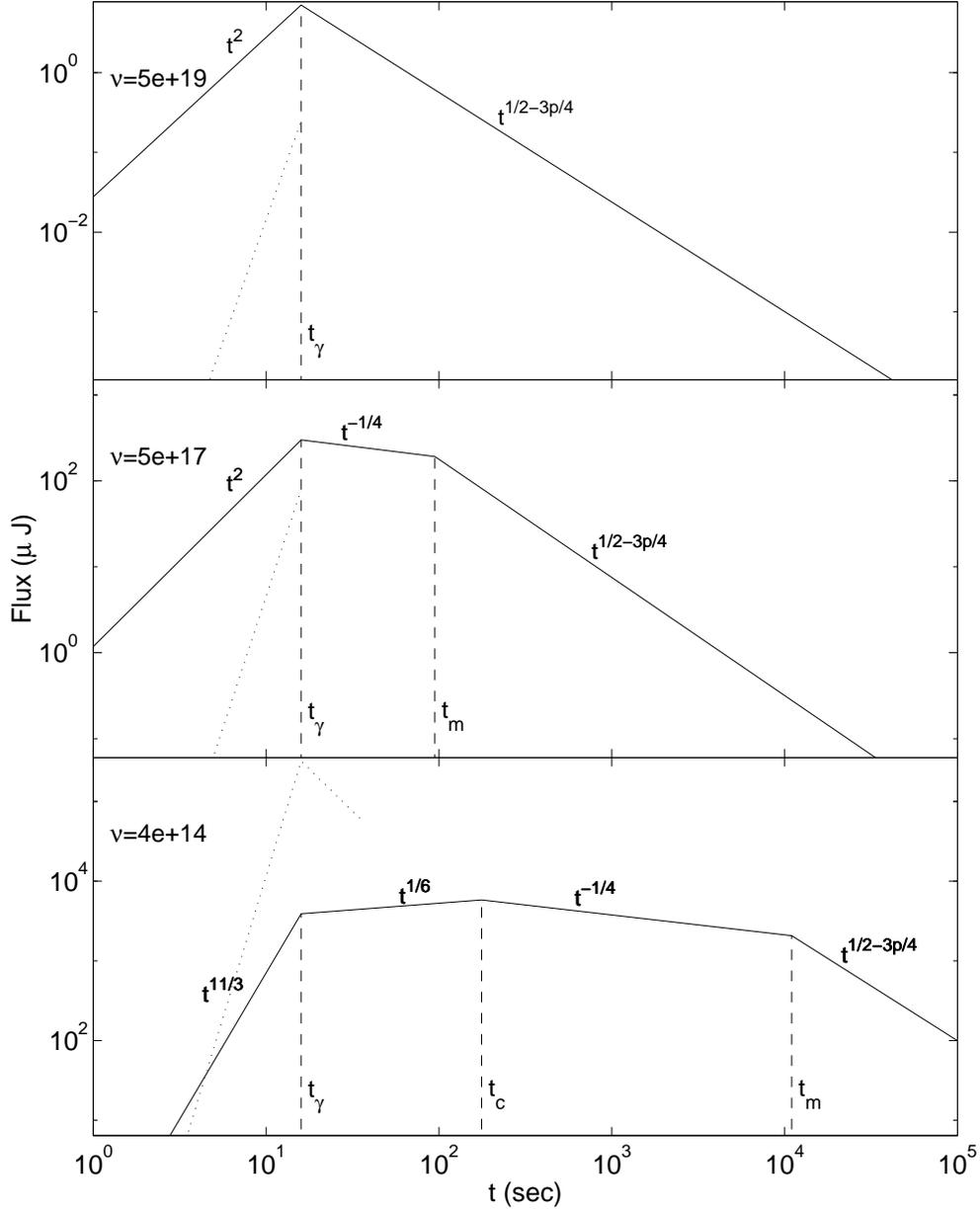}
\end{center}
\caption{The three possible light curves: high (a), intermediate (b) and low
frequencies. For high frequencies, the time $t_m$ and $t_c$ are undefined 
since $\nu_m$ is below the observed frequency throughout 
the evolution, and $\nu_c$ is crossing the observed frequency on a very
short time scale (less than a second), before fast cooling is achieved. 
In frame (b) the time $t_c<1$s therefore not seen in the plot. The dashed
line is the emission from the reverse shock. This emission terminates once
the reverse shock crosses the shell and the cooling frequency drops below
the observed frequency.}
\label{fig2}
\end{figure}

The light curve for very low frequencies, which  are typically in
the optical and possibly the UV, are  shown in figure 2c. Low
frequencies are defined by the condition: 
$t_{\gamma }<t_{c}<t_{m}$ or:
\begin{equation}
\label{lowf}
\nu <2.7\times 10^{15}\left( \frac{\epsilon _{B}}{0.1}\right) ^{-3/2}(\frac{%
\gamma _{0}}{300})^{-4/3}n_{1}^{-5/6}.
\end{equation}
At these low frequencies the 
transition to the decelerating phase occurs before the
cooling frequency crosses the observed band and before the synchrotron
typical frequency cross the observed band. Although it might be hard
to discriminate between the temporal behavior of the almost constant
($t^{1/6}$ ) and $t^{-1/4}$ parts of the low frequency light curves,
the spectral shape is very different in these two segments. The
spectrum behaves like $\nu ^{1/3}$ during the $t^{1/6}$ phase
($t_{\gamma }<t<t_{c}$) while it goes like $\nu ^{-1/2}$ during the
$t^{-1/4}$ phase ($t_{c}<t<t_{m}$). This spectral change at the time
$t_{c}$ should be sufficient to distinguish between the two segments.

For intermediated frequencies, the deceleration begins while the observed
band is above the cooling frequency but below the typical synchrotron
frequency, i.e. $t_{c}<t_{\gamma }<t_{m}$. The light curve is shown in 
figure 2b. The relevant range of frequencies are probably in the UV and or
soft X-rays, and it is intermediate between those given by the expression 
\ref{highf} and \ref{lowf}.

\section{Determination of the initial Lorentz factor}

For a short GRB, the time of the afterglow's peak is given by
equation \ref{tofgamma}. 
One can invert that to obtain the initial 
Lorentz factor   from an observed time delay: 
\begin{equation}
\gamma _{0}=\left( \frac{3E}{32\pi nm_{p}c^{5}T_{0}^{3}}\right)
^{1/8}=240E_{52}^{1/8}n_{1}^{1/8}\left( \frac{T}{10{\rm s}}\right) ^{-3/8}.
\label{gammaoft}
\end{equation}
This determination of $\gamma _{0}$ depends only  the  hydrodynamic
transition.  Therefore, it is independent of the highly uncertain
equipartition parameters $\epsilon _{B}$ and $\epsilon _{e}$ which
appear when estimating the spectrum.  It is also rather insensitive to
$E_{52}$ and $n_{1}$.  Moreover, these last two parameters can be
determined from late stage observations of the afterglow to within an
order of magnitude (Waxman 1997, Wijers and Galama 1998, Granot, Piran
and Sari 1998). This equation for $\gamma _{0}$ was used earlier, when
it was suggested that GRBs results from an external shock, to estimate
the duration of the burst (Rees \& M\'esz\'aros 1992).  However here
we assume that the external shocks produce the afterglow while
internal shocks produce the burst.

This method, for estimating the initial Lorentz factor depends on 
the identification of the  
delayed emission as resulting from the  the afterglow rather
than just another peak which is part of the burst. It is, therefore,
necessary to compare the detailed structure of the delayed emission
with the one described in the previous section. A clear characteristic
of the early afterglow emissions, in all frequencies, is an initial
very steep rise of the emission ($t^{2}$ or even $t^{11/3}$).  This
happens as the shell collects more and more material and the
interaction between the shell and the ISM becomes more and more
effective. For the thin shell light curves, discussed in this paper, this
initial rise ends at the time $t_{\gamma }$ when the deceleration
phase begins. After this rapid rise the light curve becomes almost
flat in low and intermediate frequencies, and it decreases rapidly at
high frequencies.

It is not clear if such an initial afterglow rise has been
observed so far. A good candidate might be GRB970228
(Frontera {\it et. al.} 1997, Vietri 1997). This burst consisted of a
second peak, mostly in the X-ray range. The statistics for this
event may not be good enough to enable a detailed comparison of the
light curves around its peak with the theory. However a circumstantial
evidence in favor of this explanation exists as the late time X-ray
afterglow, extrapolated back in time to the epoch of this second peak,
gives the correct flux. Note that a similar situation also exists in
GRB970508 and GRB980329, but there the flux does not seem to rise
before the second ``peak", so that the rise of the afterglow was not
observed. These bursts are probably examples of thick shells.

The identification of the second peak of GRB970228, which occurred $\sim
35$s after the burst, as the afterglow rise yields $\gamma _{0}=150$.
The estimated uncertainty is about 50\%. This arises from to the unknown
values of the energy and the external density, and from the
approximations used in the derivation of equations \ref {tofgamma} and
\ref{gammaoft} .

\section{Optical Emission and The Reverse Shock}

The fluence of a moderately strong burst is $\sim 10^{-5}$
ergs/cm$^{2}$. About one out of 5 of the BATSE bursts are stronger
than that, so such a burst occurs once a weak. Were this huge fluence
peaking at the optical band rather than in $\gamma $-Rays, with
duration of $10$sec, it would correspond to a very bright optical
source of flux
\[
\frac 1 4 \times
\frac{10^{-5}{\rm ergs/cm}^{2}}{10{\rm s} \times 5\times 10^{14}{\rm
Hz}}=50{\rm Jy}\cong 5{\rm th\;magnitude}.
\]
The additional factor of 4 in the denominator was chosen to account
for the large amount of emission above the peak frequency, which on an
average GRB goes as $F_\nu \sim \nu ^{-1.25}.$ 
The reason for taking a duration
of $10$sec is double: first it is the typical duration of a
GRB. Second it is the integration time of fast optical experiments
(LOTIS, TAROT), so that even if the emission takes place on a shorter
time scale, the effective time will be the observation's integration
time of $10$sec. However, if the emission is spread on a longer time
scale, $t_A$, then the apparent magnitude will increase accordingly by
$2.5\log _{10}(t_{A}/10sec)$.

This is by far stronger than current observational
upper limits. In fact, even a small fraction of this will be easily
observed. It is therefore worthwhile to explore the expected optical
emission at the early GRB evolution.

There are three possible emission regimes which have a comparable
amount of energy and could, in principle, emit a powerful optical
burst: the GRB itself (whether it is internal or external shocks), 
the early afterglow produced by the forward
shock, and the early emission of the reverse shock. 
Although, at their peak, each of these sights contains an
energy comparable to the total system energy, the optical signal
it produces might be  dimmer than 5th magnitude
for several reasons. The first, as we already mentioned is
if the emission is spread in time over a duration longer than $10$s. This is
simple to account for and it will increase the magnitude by 
$2.5\log _{10}(t_{A}/10sec)$.
Second, the cooling time might be longer than the system dynamical time
and the radiation is not very effective. Third, the typical emission 
frequency might peak in a different energy band rather than in the optical.
The residual optical emission might be significantly smaller. We discuss
in the following these two latter effects, and leave farther effects such
as inverse Compton scattering and self absorption to the next section.

In contrast to the previous sections, we consider here both the fast 
cooling and slow cooling synchrotron spectra given
by Sari, Piran and Narayan (1998). Ignoring self absorption, there are
four (actually five) different cases, which depend on the order of
$\nu_m$, $\nu_c$ and $\nu_{op}$ where the third frequency is a
fiducial frequency in the optical band. The fraction of the system's energy 
that is emitted in the optical band in those four cases is shown 
in Table 1.
\begin{center}
\begin{table*}[ht!]
\begin{center}
\begin{tabular}{|c||c|c|}
\hline
& $\nu_c>\nu_{op}$ & $\nu_c<\nu_{op}$ \\ \hline\hline
$\nu_m>\nu_{op}$ & $\left( \frac {\min(\nu_c,\nu_m)} {\max(\nu_c,\nu_m)}
\right) ^{1/2} \left( \frac {\nu_{op}} {\min(\nu_c,\nu_m)} \right)^{4/3^{%
\phantom{1}}}_{\phantom{1_1}} $ & $\left( \frac {\nu_{op}} {\nu_m}
\right)^{1/2}$ \\ \hline
$\nu_m<\nu_{op}$ & $\left( \frac {\nu_{op}} {\nu_m} \right)^{-(p-2)/2}
\left( \frac {\nu_{op}} {\nu_c} \right)^{1/2}$ & $\left( \frac {\nu_{op}}
{\nu_m} \right)^{-(p-2)/2^{\phantom{1}}} _{\phantom{1_1}}$ \\ \hline
\end{tabular}
\end{center}
\par
\label{t:afterglow}
\caption{The faction of the energy that get emitted 
in the optical frequency $\nu_{op}$, as function of the cooling 
frequency $\nu_c$ and the typical frequency $\nu_m$.}
\end{table*}
\end{center}

\begin{figure}[tbp]
\begin{center}
\epsscale{1.} \plotone{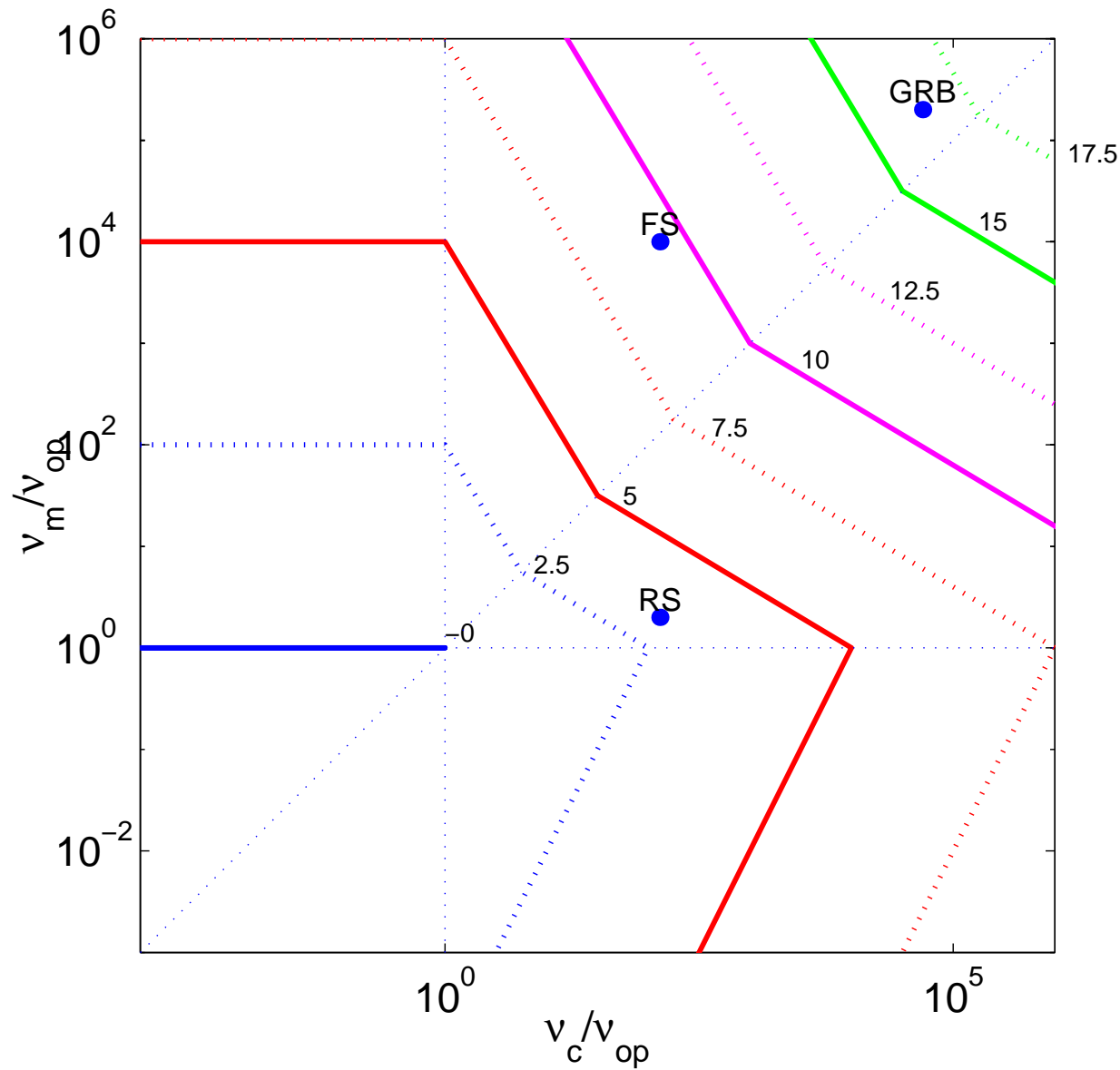}
\end{center}
\caption{ The increase in observed magnitude due to the fact that
the electrons are not cooling efficiently and/or the emission does not
peak in the optical bands. The three dots represents possible locations
of the GRB (GRB), the forward shock (FS) and the reverse shock (RS) in the
($\nu_c,\nu_m$) plane. These locations depend strongly on 
unknown parameters such as $\gamma_0$.}
\label{fig3}
\end{figure}
The corresponding increase in the magnitude is shown in
Fig. \ref{fig3}, where we have used the ``canonical'' value of $p\sim
2.5$. With this value of $p$ there is a lot of energy in the high
energy tail of the electrons distribution. Consequently the optical
emission is rather strong if $\nu _{c}<\nu _{op}$ and $\nu _{m}<\nu
_{op}$. Significant suppression occurs only if $\nu _{c}\gg \nu _{op}$
and/or $\nu _{m}\gg \nu _{op}$ with the strongest suppression taking
place if both $\nu _{m}\sim \nu _{c}\gg \nu _{op}$.

\subsection{The Prompt Optical Burst from the GRB and the Forward Shock}

Before turning to
the reverse shock, which is our main concern here we examine briefly
the prompt optical emission from the GRB and from the forward shock.
In both cases the typical synchrotron emission is
sufficiently above the optical band and hence we don't expect
significant optical emission from there.

For the GRB we use the 
observed values of $\nu_c$ and $\nu_m$. The typical emission
frequency during a GRB are mostly between 100keV and 400keV. We will
 adopt $\nu_m=5 \times 10^{19}$Hz as a typical value.
If the cooling break was below the BATSE band, then the spectral slope
within the BATSE band, independent of the electron distribution, would
have been $\le -1/2$.  The observed low energy tail is usually in the
range of -1/2 to 1/3 (Cohen et. al, 1997). This
 indicates that the cooling frequency is
close to the lower energy of the BATSE band. If the cooling break is
indeed in the BATSE range, say at $\sim 30$keV, we can substitute $\nu
_{c}=7\times 10^{18}$Hz together with $\nu _{m}=5\times 10^{19}$Hz in
table 1, to get that the residual optical emission is of $21$st
magnitude.  The point that corresponds to these parameters is marked
in Fig. 1 as GRB.  For bursts which show a low energy tail of spectral
index $-1/2$, the cooling frequency is only known to be below the
BATSE band. If, on the unlikely extreme, the cooling frequency is much
lower, say below the optical band then the optical emission is of
$11$th magnitude.

The initial emission of the forward shock, is also characterized by
very high typical synchrotron frequency and high cooling frequency (see
equations \ref{numf} and \ref{nucf}). With reasonable parameters
(e.g. $\gamma _{A}\sim 300$, $\epsilon _{e}\sim 0.5$, $\epsilon
_{B}\sim 0.1$), this emission is in the MeV range. Consequently the
optical emission is fairly weak.
Since the same forward shock is also producing
the late afterglow, one can scale late time observations to the early
epoch to obtain a direct estimate of the early value of $\nu_{c}$ and
$\nu _{m}$. Observations carried on GRB 970508 after 12 days show that
$ \nu _{m,12d} \sim 10^{11}$ Hz and $\nu _{c,12d} \sim 10^{14}$ Hz 
(Galama et. al. 1998). With adiabatic evolution $\nu _{m}\sim t^{-3/2}$ and
$\nu _{c}\sim t^{-1/2}$ so that within $10$s we expect to have $\nu
_{m}=3\times 10^{18}$ Hz and $\nu _{c}=5\times 10^{16}$ Hz. With these
values we expect the optical emission from the initial forward shock
to be of about $15$th magnitude. The point corresponding for these value
is marked on Fig 1 as FS. There is some uncertainty in this
extrapolation as the initial evolution might be radiative rather than
adiabatic. This is considerable only if the value of $\epsilon
_{e}\sim 1$ (Sari 1997, Cohen, Piran and Sari 1998). If the evolution
is initially radiative, the extrapolation according to the adiabatic
scalings is over predicting $\nu_c$ while under predicting $\nu_m$
(Sari, Piran and Narayan 1998).

\subsection{The Reverse Shock Optical Emission}

The best candidate to produce a strong optical flash is the reverse
shock (Sari \& Piran 1999). 
This shock, which heats up the shell's matter, operates only
once. It crosses the shell and accelerates its electrons. Then these
electron cool radiatively and adiabatically and settle down into a
part of the Blandford-McKee solution that determines the late profile
of the shell and the ISM. Thus, unlike the forward shock emission that
continues later at lower energies, this reverse shock emits a single
burst with a duration comparable to $t_{A}$ (the duration of the GRBs
or a few tens of seconds if the burst is short). 
After the peak of the
reverse shock, i.e. after the reverse shock has crossed the shell no
new electrons are injected. Consequently there will be no longer
emission above $\nu _{c}$, and $\nu _{c}$ drops fast with time due to
adiabatic cooling of the shocked 
shell material. Therefore, in contrast
to the forward shock were we have calculated the whole light curve,
we will focus here on the emission at the peak time $t_A$.

This peak time is given by 
\begin{equation}
t_{A}={\rm \max }[t_\gamma, \Delta /c]
\end{equation}
and the Lorentz factor at this time is
\begin{equation}
\gamma_A={\rm \min}[\gamma_0,(17E/128\pi\Delta^3 n m_p c^2)^{1/8}]
\end{equation}
The afterglow typical time is similar to the duration of the burst
(if the burst is long or the initial Lorentz factor is large) or longer than
that if the burst was short and the initial Lorentz factor was low. In the
latter case the shells Lorentz factor at the time $t_{A}$ equals its
initially value $\gamma =\gamma _{0}$, while in the former case some
deceleration has already occurred and $\gamma <\gamma _{0}$. After this
time, $t_{A}$, a self similar evolution begins, and the initial width of the
shell is no longer important. Therefore, the Lorentz factor at the time $%
t_{A}$, could be estimated in both cases as
\[
\gamma _{A}=\left( \frac{17E}{128\pi n m_p c^5 t_A}\right) ^{3/8}.
\]

Before discussing the details of this emission we outline a simple
energetic consideration to show that the initial energy dissipated in
the reverse shock is comparable to the initial energy dissipated in
the forward shock (Sari and Piran 1995) and to the GRB energy. The
forwards shock and the reverse shock are separated by a contact
discontinuity, across which there is a pressure equality.  This means
that the energy density in both shocked regions is the same. As the
forward shocks compresses the fluid ahead of it by a factor of $\gamma
^{2}$ its width is of order $R/\gamma ^{2}$. Though the initial width
of the shell can be smaller than that, it will naturally spread to
this size due to mildly relativistic expansion in its own frame. Since
the energy density is the same and the volume is comparable, the total
energy in both shocks is comparable. A more detailed calculation (Sari
and Piran 1995) shows that at the time the reverse shock crosses the
shell about half of the energy is in the shocked shell material.

The two frequencies that determine the spectrum, $\nu _{c}$ and
$\nu _{m}$ for the reverse shock are most easily calculated by
comparing them to those of the forward shock.  The equality of energy
density across the contact discontinuity suggests that the magnetic
fields in both shocked regions are the same (provided, of course, that
we assume the same magnetic equipartition parameter in both regions).
Both shocked material move with the same Lorentz factor.  Therefor,
the cooling frequency, $\nu _{c}$, at the reverse shock is equal to
that of the forward shock. 
However, instead of using the general description
of this frequency as function of both $\gamma$ and $t$, we can substitute
the expression for $\gamma_A$ to get:
\begin{equation}
\nu _{c}=8.8\times 10^{15} {\rm Hz}\left( \frac{\epsilon _{B}}{0.1}\right)
^{-3/2}E_{52}^{-1/2}n_{1}^{-1} t_A^{-1/2}
\end{equation}

The typical synchrotron frequency is proportional to the electrons
random Lorentz factor squared (temperature square) and to the magnetic
field and to the Lorentz boost. This leads to the $\gamma ^{4}$
dependence in equation \ref{numf}. The Lorentz boost and the magnetic
field are the same for the reverse and forward shocks while the random
Lorentz factor is $\gamma _{0}/\gamma _{A}$ compared to $\gamma _{A}$
of the forward shock.  The ``effective'' temperature at the reverse
shock is much lower than that of the forward shock (by a factor of
$\gamma _A^2/\gamma_0 \gg 1$).  The reverse shock frequency at the
time $t_A$ is, therefore, given by:
\begin{equation}
\nu _{m}=1.2\times 10^{14}\left( \frac{\epsilon _{e}}{0.1}\right)
^{2}\left( \frac{\epsilon _{B}}{0.1}\right) ^{1/2}(\frac{\gamma _{0}}{300}%
)^{2}n_{1}^{1/2}.
\end{equation}
So while the forward shock radiates initially at the energies
of $\sim $MeV the reverse shock radiates at few eV, with
significant radiation emission within the optical band.

The case, which is most favorable for a strong optical emission is if
the typical frequency of the reverse shock falls just in the optical
regime and if the cooling frequency is on or below the optical
frequency. This can be achieved with reasonable parameters, say with
$n_{1}=E_{52}=1$, $\epsilon _{B}=0.2$, $\epsilon _{e}=0.5$ and $\gamma
_{0}=100$. The other extreme case, which has a considerable lower
optical fluence is if the typical radiation frequency
as well as the cooling frequency are above the optical regime. As is
apparent from these last two equations this requires a high initial
Lorentz factor, short GRB, high electron equipartition parameter and a low
magnetic equipartition parameter.

The resulting optical emission, as function of the most unknown variable 
$\gamma_0$, and for the ``best guess" value of the other parameters, as obtained
by late afterglow observations, is given in figure 3. As the Lorentz 
factor increases the optical emission initially rises. This is mainly due 
to the fact that the emission is spread on a shorter time scale
($t_A$ is decreasing).
However, with quite a moderate initial Lorentz factor ($\gamma_0 \sim 300$)
the emission
duration does no longer depend on the initial Lorentz factor but is given 
by the observation's integration time (which we assumed to be $10$sec) 
or by the duration of the burst (for bursts longer than $10$sec). 
As the Lorentz factor continues to increase,
the emission drops due to the increase in $\nu_m$. 
With high enough values of $\gamma _{0}$ the flux decreases considerably. 

\begin{figure}[tbp]
\begin{center}
\epsscale{1.} \plotone{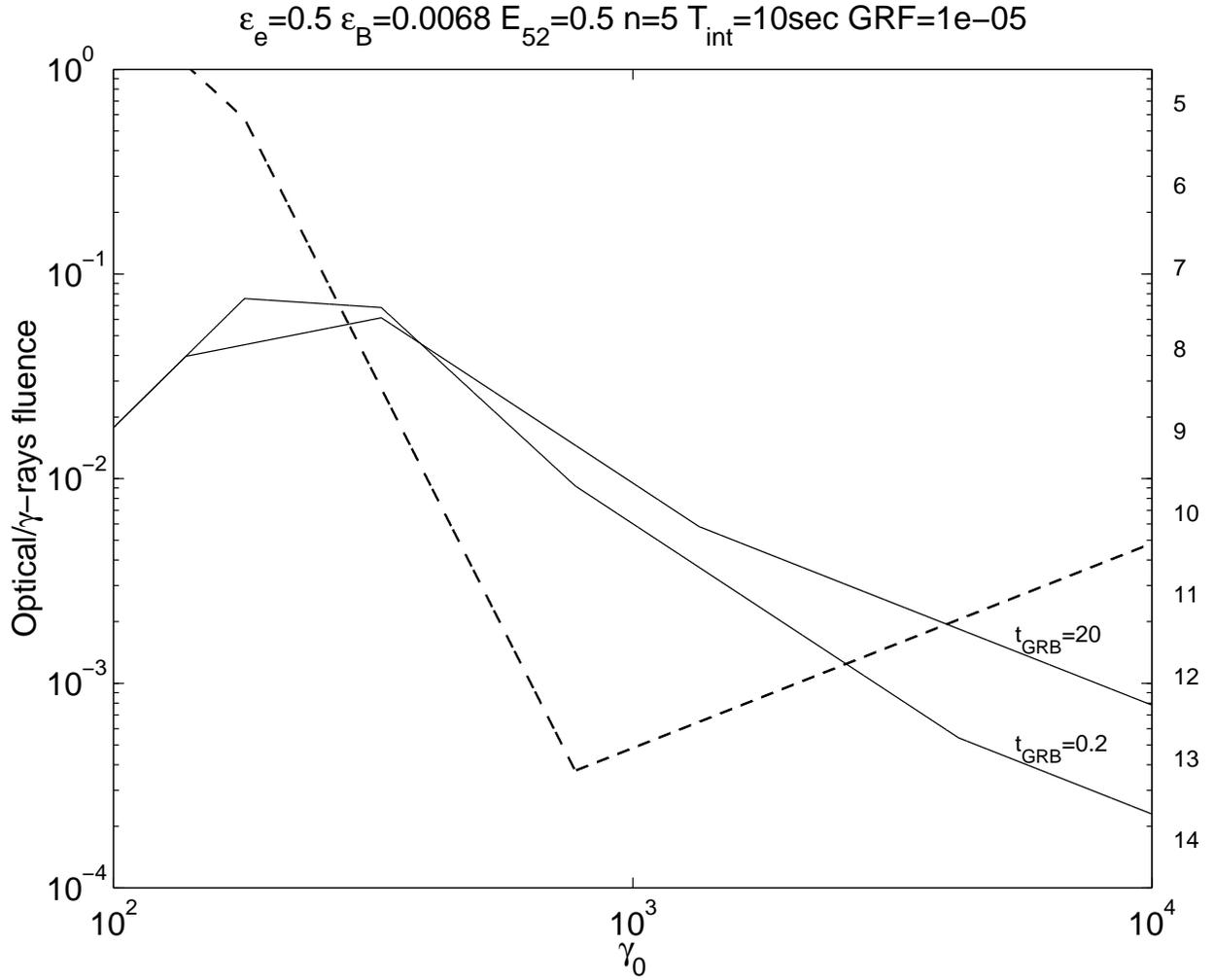}
\end{center}
\caption{ The optical flash from the reverse shock at its peak. The dashed
line is the maximal emission allowed by self absorption for burst of
duration 0.1s. The self absorption limit for the long burst is too high to
fit the y axis.}
\label{fig4}
\end{figure}

Two other effect can reduce the flux below these estimates: Self
absorption might reduce this flux if the system is optically thick at
optical frequencies; Inverse Compton may compete with synchrotron in
cooling the electrons and reduce the synchrotron flux. We consider
these effects now.

\subsection{Synchrotron Self Absorption}

Self absorption would reduce the optical flux from the reverse shock
if it is optically thick.  A simple way to account for this effect, is
to estimate the maximal flux emitted as a black body with the reverse
shock temperature.  This is given by the
\begin{equation}
F_{sa}=\pi (R_{\perp }/D)^{2}S_{\nu }=\pi \left( \frac{R_{\perp }}{D}\right)
^{2}\frac{2\nu ^{2}}{c^{2}}\frac{\epsilon _{e}}{3}m_{p}c^{2}\gamma _{0},
\end{equation}
where the quantity $R_{\perp }\sim \gamma _{A}ct_{A}$ is the observed size
of the fireball. More detailed calculations,
(Waxman 1997, Panaitescu \& M\'esz\'aros 1997, 
Sari 1998, Granot, Piran and Sari 1998a,b) obtain a size bigger by a 
factor of $\sim 2$. However, these are applicable only deep inside the 
self-similar deceleration while we are interested in its beginning.
To be conservative, we use the lower estimate of the size, which result
in a weaker emission. We get that
\begin{equation}
\label{sa}
F_{sa}=4.8{\rm Jy} D_{28}^{-2} E_{52}^{1/4}n_1^{-1/4}
\frac {\epsilon_e} {0.1}
\frac {\gamma_0} {300}
\left( \frac {t_A} {1s} \right)^{5/4}
\end{equation}
Note that so far we have eliminated the dependence on the distance to
the burst $D$ by using the observed fluence of the burst. However,
self absorption depend on the flux per unit area at the source. The
distance, therefore, appears explicitly and can not be
eliminated. With a given observed flux, the further the burst is the
more important is self absorption.
It can be seen from equation \ref{sa} that self absorption can hardly play any
role for long bursts with say $t_A>1$sec.  Self absorption can be important
only if $t_A$ is very short, which is possible only for short bursts
and high values of $\gamma_0$. 

\subsection{Inverse Compton Cooling}

Synchrotron self-Compton, that is inverse Compton scattering of the
synchrotron radiation by the hot electrons provides an alternative way
to cool the electrons. The typical energy of a photon that has been
scattered is $\nu _{IC}\sim \nu _{m}\gamma _{e}^{2}$. This emission
will be in the MeV regime and not in the optical band. However, 
if $\epsilon_e>\epsilon_B$ then the
efficiency of inverse Compton as a cooling mechanism, relative to
synchrotron emission equals: $ \sqrt{\epsilon _{e}/\epsilon _{B}}$
(Sari, Narayan and Piran 1996). This will reduce the synchrotron flux of
any cooling electron by that factor but will not alter the emitted flux of 
a non cooling electron. It will therefore influence the optical emission only
if $\nu_c<\nu_{op}$, and may reduce the flux for this case by a factor of
few, resulting in the increase of one or two magnitudes. However, 
if $\nu_c<\nu_{op}$ the reverse shock synchrotron flux is very high 
to begin with.

\subsection{Extinction}

In all the discussion above, we have normalized the optical flux according 
to the observed GRB fluence. However, $\gamma$-rays do not suffer any kind 
of extinction, while the optical regime may do. Some afterglows show only
small amount of extinction, some show strong extinction while other 
do not show any optical activity and are speculated to be in a highly 
extincting surrounding. Extinction is probably important if the burst is 
located in a star formation regime.
GRB970508, for example, shows only week extinction after its peak
at 2 days. However, before this peak the optical light curve does not fit
any of the prediction of the simple models. 
If this is due to extinction that disappears after
two days, it might be crucial in the first few seconds in which we 
are interested.

\section{Discussion}

We have calculated the observed synchrotron spectra expected from a
relativistic shock that accelerates electrons to a power law
distribution for an arbitrary hydrodynamic evolution $\gamma (t)$. 
Light curves can be obtained from
this spectra by substituting initially $\gamma (t)=conts$, then $
\gamma (t)=t^{-1/4}$ and finally $\gamma (R)=t^{-3/8}$.
Where the intermediate expression relevant only for thick shells.  For
thin shells, we have explicitly constructed the possible light curves
for the forward shock,
for several frequency regimes.  We find that the flux must rise
initially steeply as $t^{2}$ or as $ t^{11/3}$. This rapid rise ends
at the time $t_\gamma$ when the system approaches self similar
deceleration. After this time the light curve is either decreasing
(high frequencies) or almost flat (low and intermediate
frequencies). The break at $t_\gamma$ is, therefore, quite sharp and
an observational determination of this transition time should be
simple.

In the internal-external scenario, thin shells corresponds to short
bursts. We expect, in this case, a gap between the burst and its
afterglow. This gap allows a clean observation of the early afterglow
light curve. In particular we should observe a clear rise, which is
not contaminated by the complex, variable internal shocks burst.
Thick shells (which correspond to longer bursts) 
light curves are more complex, due to the overlap of the
burst and the afterglow. This overlap would make it difficult or even
impossible to isolate the early afterglow signal.

A detection of the early afterglow rise is possible with future
missions.  Observations of this predicted light curves would confirm
the internal shocks scenario. They will also enable us to measure the
initial Lorentz factor. Both these ingredients are essential in order
to build a reliable source model. As long as the question of internal
or external shocks is not settled with a high certainty, it is not
clear whether the source deriving the whole phenomena is operating for
a millisecond (as required to the fireball needed for the external
shock scenario) or for hundred seconds (as required for the internal
shocks).

It is important to stress that a detection of a gap in the emission,
by it self, is meaningless. The later emission that follows the gap
can be just another peak in the complex GRB emission produced by the
internal shocks.  A comparison between this emission and the
theoretical prediction given here is needed in order to unambiguously
identify any delayed emission as the beginning of the afterglow rather
than as a continuation of the burst. The spectra and light curves
described here should be used to discriminate between an additional
``delayed'' peak, which is just a part of the internal shock burst and
an emission coming from an external shock which should be described by
the smooth light curves given here.

A broad band detection of the spectrum (say at 1-1000keV), at the time
that the afterglow peaks, will enable us to compare between the
spectral properties of the GRB and those of the afterglow. In the
internal-external picture these spectra are not closely related and
the typical synchrotron frequency and cooling frequency emitted in the
early afterglow can be either higher or lower than that of the burst.
On the other hand, the burst and the afterglow should be similar if
the burst itself is also produced by external shocks.

We have calculated the optical emission that is expected in the simplest
scenario of creating GRBs. The emission in the optical regime is dominated
by the reverse shock. We showed that a strong optical flash is 
expected over a duration comparable to that of the GRB or delayed a few
dozens of seconds after that.
We have used the terminology of the internal-external
scenario, where the GRB is produced by internal shocks while the afterglow
by external shocks. However, even if the GRB is also produced 
by external shocks, our conclusions are still valid, with $t_A$ being the
duration of the GRBs itself. The problem in this case, is that the assumption
of a uniform surrounding may not be valid for models producing the GRB
by external shocks.

The calculations regarding the reverse shock emission assumed that the 
shell is made out of baryonic material. If instead it is magnetically 
dominated where the energy in 
the rest mass of the baryons is negligible, a considerably lower emission
is expected from the reverse shock.
Our prediction is heavily based on the fact 
that the reverse and forward 
shock carry the same amount of thermal energy. If the shell is initially
very thin, and somehow does not spread so that its thickness is kept 
significantly below $R/\gamma^2$, the reverse shock will be Newtonian, and 
will contain a small fraction of the system energy. The emission will 
be reduced accordingly. In the simplest model, where the shell was accelerated
hydrodynamically, the back of the shell moves with a Lorentz factor smaller
by a factor of a few from its front (this is what defines the shell) so that
spearing to a thickness of $R/\gamma^2$ is unavoidable. However, in
more complicated forms of acceleration, one might think of shells that
have a perfectly uniform Lorentz factor and therefore does not spread.

If the density of the surrounding material is very low, it might take a
long time before the shell begins to decelerate, i.e. a very large $t_A$. 
The reverse shock emission
will be spread over this large time, resulting in a much lower magnitude. 
However, it seems to be that this possibility of long $t_{A}$ is already
ruled out by current observations, as the beginning of the X-ray decay was 
observed with BeppoSAX just following some bursts, like GRB 970228, 
GRB970508 and GRB971214.

Fast optical followup experiments often have a tradeoff between the magnitude
they can achieve and how fast can they operate. In this respect, an optical
experiment which can detect emission which is simultaneous with the burst
is preferred since the reverse shock emission might die soon after that.
As there are many bursts of duration of 10 seconds or above, this might
be the optimal response time for an optical follow up. Nevertheless,
experiments with delays of $30-100$s should still be able to detect the
reverse shock emission from a few long enough bursts.

Finally there is the possibility of extinction. At least in some burst, like
GRB 970508, extinction does not seem to play a very important roll in 
the late afterglow. 
However, the early signal of GRB 970508 (before its peak at two days) 
is not described well by the theory. If this is an evidence of some 
extinction, which is important only on early times, it might reduce 
the optical flash predicted here.

\acknowledgments
This research was supported by the US-Israel BSF grant 95-328, by a
grant from the Israeli Space Agency and by NASA grant NAG5-3516.
Re'em Sari thanks the Sherman Fairchild Foundation for support. Tsvi
Piran thanks Columbia University and Marc Kamionkowski for hospitality
while this research was done.

\end{document}